\documentclass[%
reprint,
superscriptaddress,
amsmath,amssymb,
aps,
]{revtex4-1}

\usepackage{graphicx}
\usepackage{dcolumn}
\usepackage{bm}
\usepackage{comment}
\usepackage{hyperref}

\usepackage{makecell}
\usepackage{changepage}
\usepackage{multirow}
\usepackage[table]{xcolor}
\usepackage{cleveref}
\newcommand{\beginsupplement}{%
\setcounter{table}{0}
\renewcommand{\thetable}{S\arabic{table}}%
\setcounter{figure}{0}
\renewcommand{\thefigure}{S\arabic{figure}}%
\renewcommand{\theHtable}{Supplement.\thetable}
\renewcommand{\theHfigure}{Supplement.\thefigure}
}

\begin{document}

\title{SEIRD model to study the asymptomatic growth during COVID-19 pandemic in India}

\author{Saptarshi Chatterjee}
\email{sspsc6@iacs.res.in}
\affiliation{Indian Association for the Cultivation of Science, Kolkata-700032, India}

\author{Apurba Sarkar}
\email{sspas2@iacs.res.in}
\affiliation{Indian Association for the Cultivation of Science, Kolkata-700032, India}

\author{Mintu Karmakar}
\email{mcsmk2057@iacs.res.in}
\affiliation{Indian Association for the Cultivation of Science, Kolkata-700032, India}

\author{Swarnajit Chatterjee}
\email{sspsc5@iacs.res.in}
\affiliation{Indian Association for the Cultivation of Science, Kolkata-700032, India}

\author{Raja Paul}%
\email{raja.paul@iacs.res.in}
\affiliation{Indian Association for the Cultivation of Science, Kolkata-700032, India}%

\begin{abstract}
According to the current perception, symptomatic, presymptomatic, and asymptomatic infectious persons can infect the healthy population susceptible to the SARS-Cov-2. More importantly, various reports indicate that the number of asymptomatic cases can be several-fold higher than the reported symptomatic cases. In this article, we take the reported cases in India and various states within the country till September 1, as the specimen to understand the progression of the COVID-19. Employing a modified SEIRD model, we predict the spread of COVID-19 by the symptomatic as well as asymptomatic infectious population. Considering reported infection primarily due to symptomatic we compare the model predicted results with the available data to estimate the dynamics of the asymptomatically infected population. Our data indicate that in the absence of the asymptomatic infectious population, the number of symptomatic cases would have been much less. Therefore, the current progress of the symptomatic infection can be reduced by quarantining the asymptomatically infectious population via extensive or random testing. This study is motivated strictly towards academic pursuit; this theoretical investigation is not meant for influencing policy decisions or public health practices.
\end{abstract}

\maketitle


\section*{Introduction}
\begin{figure}[htbp!]
\includegraphics[width=\linewidth]{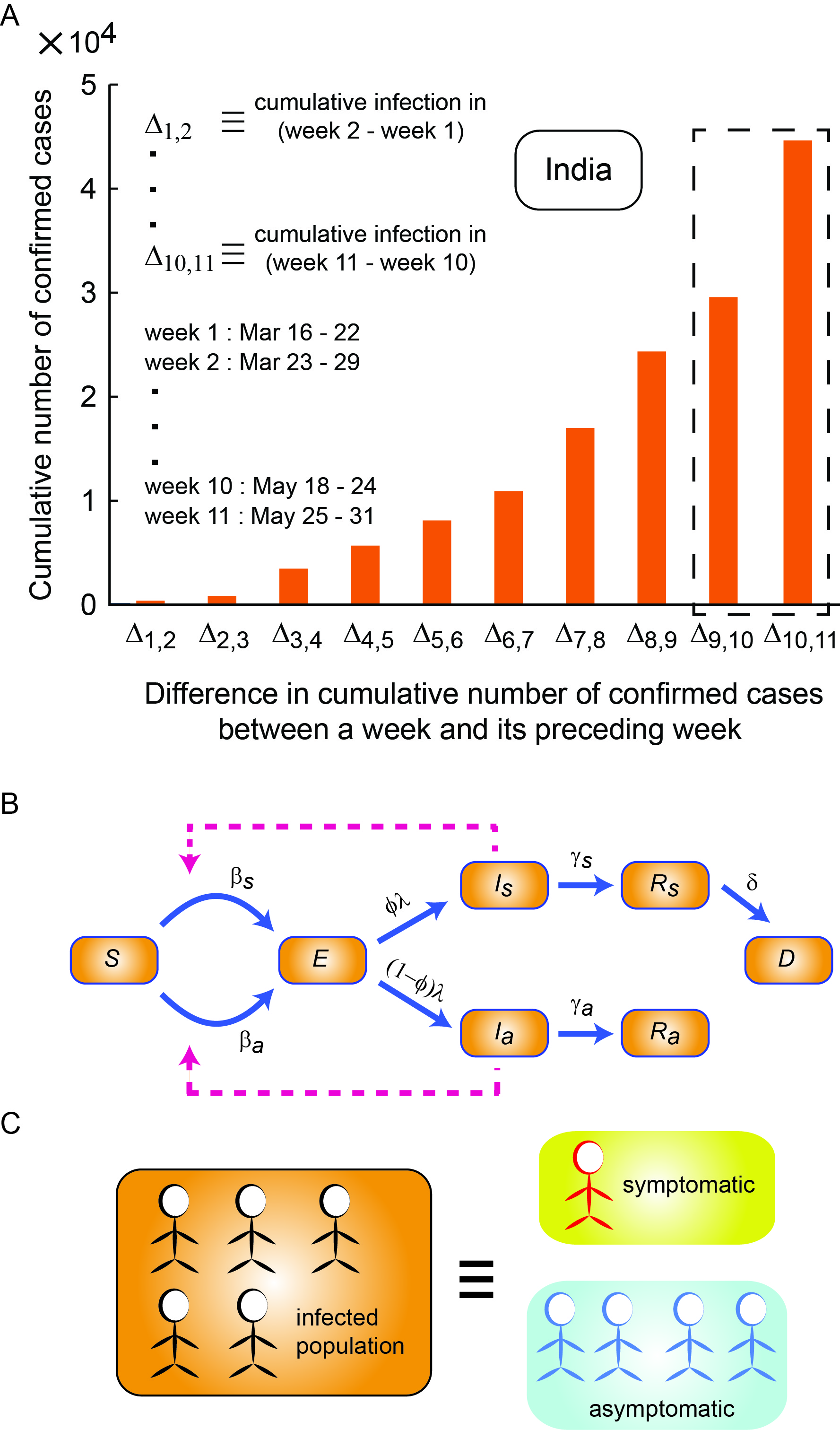}
\caption{ COVID-19 progression in India and the modified SEIRD model used for mapping the current COVID-19 trajectory in India's context. (A) Increment in the cumulative number of confirmed cases in India. The bars boxed in black highlight the surge of infections in the early stages of COVID-19 pandemic in India. (B) Schematic representation of modified SEIRD model. The arrows denote the flux direction between the compartmentalized sub-populations of susceptible ($S$), exposed ($E$), symptomatically infected ($I_{s}$), asymptomatically infected ($I_{a}$), recovered from symptomatic infection ($R_{s}$), recovered from asymptomatic infection ($R_{a}$) and dead ($D$). (C) The COVID-19 infected individuals can either develop symptoms or remain asymptomatic throughout before recovery. According to the reports by the WHO and the ICMR, India, the asymptomatic infections can be as large as 4-5 times of the total symptomatic infections. } \label{fig1}
\end{figure}

At the beginning of the outbreak of COVID-19, the general perception was that the transmission of the disease occurred mostly through the infectious persons having influenza-like symptoms. One reason behind this view was the similarities between the SARS pandemic in 2003 caused by SARS-CoV-1 and the current threat COVID-19 caused by SARS-CoV-2.  Both SARS and COVID-19 patients show similar influenza-like symptoms,  propagate infections through person-to-person contact via respiratory droplets generated when an infected person breathes, coughs, and sneezes. Measures taken to successfully control the SARS in 2003 were majorly based on testing those having symptoms and isolating the positive cases. Nevertheless, a similar guideline has not been very effective in controlling the COVID-19 as the number of infected individuals has gone past 4 million in India and 27 million worldwide. Despite severe containment measures, these numbers are several orders of magnitude higher than SARS in 2003 that reported 8,422 cases worldwide with a case fatality rate of 11\%. In India, even when the nation was under lockdown over the past few months, a noticeable surge in the new COVID-19 positive cases was observed (Fig. \ref{fig1}A). Initially, when the nation-wide lockdown was enforced on March 23, the infection growth showed a gradual slowdown for a few weeks. However, as days passed, the number of daily new confirmed cases increased significantly which prompted us to revisit our earlier work \cite{Chatterjee2020} on the COVID-19 outspread in India using the well-known SIRD model \cite{KM1927,KM1932,Brauer2019,Daley2009,Fanelli2020,Fernandez2020,Biswas2020,Mukherjee2020}. Normally, an individual exposed to the disease sufficient to catch the infection goes through an incubation period of several days ($\sim$ 5 days \cite{Bar-On2020}) before showing symptoms, i.e. presymptomatic. However, many reports \cite{Ferretti2020, Davies2020, Furakuwa2020, Mizumoto2020, Ye2020, Kissler2020} suggested that a large fraction of those who catch the disease do not show any symptoms at all, i.e. they remain asymptomatic (tested positive for SARS-CoV-2 RNA but show no signs of illness). The presence of the virus in presymptomatic or asymptomatic persons means that they can transmit the disease to others. In this study, we investigate whether the trajectory of the continuous upsurge of infections (Fig. \ref{fig1}A) can be attributed to the following characteristics of this disease spread: (a) in addition to the symptomatic cases, the abundance of the infected individuals who do not develop any symptoms at all or remain mildly symptomatic, before recovery (asymptomatic) and (b) human-to-human transmission via those asymptomatic carriers. We use a modified version of the well known SEIRD model in epidemiology \cite{Leon2020,Gatto2020} as a working handle to account for the asymptomatic infections (Fig. \ref{fig1}B-\ref{fig1}C). The model analysis as per the current data for India and few Indian states shows that the dynamics of the infection spreading is highly influenced by the asymptomatic population.
The real data for India and various states within the country are acquired from the repository with an interactive handle hosted at \url{https://www.covid19india.org}.
The purpose of this article is neither to influence policy decisions nor to put forward any quantitative projections that should be used to design public health guidelines.

\section*{Model}
In the standard SEIRD model, the population $N$ is divided into sub-population of susceptible ($S$), exposed ($E$), infected ($I$), recovered ($R$) and dead ($D$) for all times $t$. Thus, $N = S + E + I +R + D$. In our case, in order to account for the asymptomatic population, we introduced several new `sub-population' compartments and modified the corresponding flows between these compartments. The infected population $I$ in SEIRD model is sub-divided into two compartments: (a) infected symptomatic population $I_{s}$ and (b) infected asymptomatic population $I_{a}$ (Fig.\ref{fig1}B).
As per the WHO and the ICMR, India reports, the COVID-19 infected individuals can be divided into two sub-groups: (a) symptomatic carriers and (b) asymptomatic carriers, infected individuals without any symptoms. The asymptomatic population is reported to be several times larger in number compared to the symptomatic population (Fig. \ref{fig1}C).
Similarly, the population of recovered $R$ in the SEIRD model is also classified into two new compartments: (a) population recovered from symptomatic infection $R_{s}$ and (b) population recovered from asymptomatic infection $R_{a}$ (Fig.\ref{fig1}B). The population of dead ($D$) has inflow from $I_{s}$ only (Fig.\ref{fig1}B); we assumed no death for the asymptomatic individuals due to the disease.

The following set of mean-field differential equations dictates the time-dependent dynamics of the population of susceptible ($S$), exposed ($E$), symptomatically infected ($I_{s}$), asymptomatically infected ($I_{a}$), recovered from symptomatic infection ($R_{s}$), recovered from asymptomatic infection ($R_{a}$) and dead ($D$), comprehensively capturing the time evolution of the outspread across the total population. 

\begin{equation}
{\frac{dS(t)}{dt}=-\beta_{s} S I_{s}/N - \beta_{a} S I_{a}/N}
\label{eq1}
\end{equation}
\begin{equation}
{\frac{dE(t)}{dt}=\beta_{s} S I_{s}/N + \beta_{a} S I_{a}/N - \lambda E}
\label{eq2}
\end{equation}
\begin{equation}
{\frac{dI_{s}(t)}{dt}= \phi \lambda E - (\gamma_{s} + \delta) I_{s}}
\label{eq3}
\end{equation}
\begin{equation}
{\frac{dI_{a}(t)}{dt}= (1-\phi) \lambda E - \gamma_{a} I_{a}}
\label{eq4}
\end{equation}
\begin{equation}
{\frac{dR_{s}(t)}{dt}=\gamma_{s} I_{s}}
\label{eq5}
\end{equation}
\begin{equation}
{\frac{dR_{a}(t)}{dt}=\gamma_{a} I_{a}}
\label{eq6}
\end{equation}
\begin{equation}
{\frac{dD(t)}{dt}=\delta I_{s}}
\label{eq7}
\end{equation}

The parameters $\beta_{s}$, $\beta_{a}$, $\lambda$, $\phi$, $\gamma_{s}$, $\gamma_{a}$ and $\delta$ regulate the temporal flow between different sub-populations of infected, recovered and dead (Fig. \ref{fig1}B). Note that, in the above set of equations, the susceptible population ($S$), upon interactions with symptomatic and asymptomatic population, becomes exposed ($E$) to the disease with contact rates $\beta_{s}$ and $\beta_{a}$, respectively. The model construction alludes that the individuals belonging to the exposed population are not contagious; hence, they do not transmit the disease by infecting others. The disease transmission from one individual to another occurs when an exposed individual transforms into either a symptomatic or an asymptomatic carrier of the disease (Fig. \ref{fig1}B-\ref{fig1}C). The time-dependent out-flux from the exposed compartment ($E$) into symptomatic and asymptomatic compartments ($I_{s}$ and $I_{a}$) is governed by the rate $\lambda$. The out-flux from the exposed compartment with the rate $\lambda$, develops symptomatic infections with a probability $\phi$ and asymptomatic infection with a probability $(1-\phi)$. The recovery of a symptomatic infected person and an asymptomatic infected are determined by the rates $\gamma_{s}$ and $\gamma_{a}$, respectively. An infected individual ceases to transmit the disease once s/he is recovered. Previous reports \cite{Bar-On2020,XiHe2020,Li2020} indicate that, on the average, contagiousness begins to pronounce from 2-3 days before the appearance of symptoms. The contagiousness (disease spreading capability) of a symptomatically infected individual mounts to its peak before the symptoms develop and stays for about 7-9 days after the peak infection is passed. Therefore, a symptomatic individual bears the ability to transmit the disease for nearly about 12 days on average before recovery. However, how long it takes for an asymptomatic individual to recover is hard to ascertain, as a large number of asymptomatic infections remain undetected due to the sheer lack of symptoms. For symptomatic recovery, the time scale of $\sim$ 12 days provides an estimate of the magnitude of the symptomatic recovery rate $\gamma_{s}$. In the preliminary model analysis, the magnitude of the asymptomatic recovery rate $\gamma_{a}$ is chosen to be close to the magnitude of symptomatic recovery rate $\gamma_{s}$. Deaths ($D$) within the symptomatic population occur with a rate $\delta$. We assume that infected, yet asymptomatic individuals do not die due to this disease.
Nevertheless, it is imperative to mention that all the relevant rate parameters are varied within a reasonably feasible range to obtain the best fit of the theoretical curves with the real data. 

\subsubsection*{Estimating the spread of the disease, the reproduction number}
The basic reproduction number denoted by $R_{0}$ is a crucial parameter in epidemiology that can indicate the pandemic situation of an infectious disease. It is defined as the average number of secondary infections produced by a primary infection seeded into a sea of the susceptible population and its value is suggestive to implement a control intervention for containing the disease. However, in most of the practical situations, it is difficult to realize the single primary infection that has caused the outbreak; besides all contacts may not be susceptible to the infection. Therefore, in the present circumstances, instead of using the term basic reproduction number, we denote this by an effective reproductive number $R_{e}$. The effective reproductive number is determined by various rates presented in \Cref{eq1,eq2,eq3,eq4,eq5,eq6,eq7}. In general, an epidemic will begin when $R_{e}>1$, so the number of infections increases. Pandemic will become an endemic when $R_{e}$ becomes 1 and $R_{e}<1$ will eventually diminish the number of infections. The effective reproduction number $R_{e}$ can be computed from the \Cref{eq1,eq2,eq3,eq4,eq5,eq6,eq7} using the next generation matrix NGM~\cite{Diekmann2010}. An expression for $R_{e}$ in the pre-lockdown situation is given by: 
\begin{equation}
R_{e} = \frac{\phi \beta_{s}}{\delta + \gamma_{s}} + \frac{(1-\phi) \beta_{a}}{\gamma_{a}}    
\label{eq8}
\end{equation}
The mathematical treatment for obtaining the closed-form expression of $R_{e}$ using NGM is similar to that given in \cite{Leon2020}.

\subsubsection*{Accounting for the effects of enforced containment measures}
Before discussing the effect of containment measures on the infection dynamics, we need to highlight the subtle differences in the connotations of the following terms: (a) lockdown, (b) containment measures, and (c) social distancing. We stress that a lockdown in the model indicates the containment measures enforced across the country. But, even without a countrywide lockdown, containment can be imposed locally in the hotspots of infections (as being implemented during the current unlock phase) to prevent further spread of the disease into a larger region. In a sense, the word `lockdown' bears an essence of universality whereas the implementation of containment can be both local and universal. However, social distancing is not an external norm to be enforced. It is rather a choice of lifestyle where one maintains a `good' habit of physical distancing with other individuals in a public place/crowded environment. Ideally, even if there is no lockdown and/or containment measures in place, strict maintenance of social distancing can significantly reduce the chances of new infections. In the manuscript, in a general sense, where we mention the term `lockdown', we imply that containment measures and/or social distancing are in effect.

Under normal circumstances, the implementation of the containment measures would reduce the interaction of the susceptible population ($S$) with the infected population ($I_{s}$ and $I_{a}$). As gleaned from the \Cref{eq1}, the susceptible population, while interacting with the symptomatically and asymptomatically infected population, becomes exposed to the disease with rates $\beta_{s}$ and $\beta_{a}$, respectively. We can consider that the direct impact of the enforced containment measures (lockdown in a nation-wide sense) and social distancing would be reflected on $\beta_{s}$ and $\beta_{a}$ in \Cref{eq1,eq2}. The susceptible-to-exposed transition rates $\beta_{s}$ and $\beta_{a}$ should diminish with time as the lockdown is enforced and continued. Hence, as a simple choice, we set the susceptible-to-exposed transition rates $\bar{\beta_{s}} (t)$ and $\bar{\beta_{a}} (t)$ in such a way that these rates gradually decline with time as the containment measures are put in place \cite{Fanelli2020,Fernandez2020}. Prior to the lockdown, $\beta_{s}$ and $\beta_{a}$ are constant. During the lockdown, the time-dependent $\bar{\beta_{s}}(t)$ and $\bar{\beta_{a}}(t)$ are assumed to decrease exponentially as depicted in the following~\cite{Fanelli2020}.
During the mixing between susceptible population ($S$) and symptomatically infected population ($I_{s}$),
\begin{equation}
\bar{\beta_{s}}(t) = \beta_{s} (1 - \xi_{s}) e^{- [(t - \tau)/T]} + \xi_{s}\beta_{s} \ \text{for} \ t > \tau
\label{eq9}
\end{equation}
\begin{equation}
\bar{\beta_{s}}(t) = \beta_{s} \ \text{for} \ t \leq \tau
\label{eq10}
\end{equation} 

Similarly, During the mixing between susceptible population ($S$) and asymptomatically infected population ($I_{s}$),
\begin{equation}
\bar{\beta_{a}}(t) = \beta_{a} (1 - \xi_{a}) e^{- [(t - \tau)/T]} + \xi_{a}\beta_{a} \ \text{for} \ t > \tau
\label{eq11}
\end{equation}
\begin{equation}
\bar{\beta_{a}}(t) = \beta_{a} \ \text{for} \ t \leq \tau
\label{eq12}
\end{equation} 

Here, the lockdown (enforcement of the containment measures) begins on the day $\tau$ which is calculated from the day $0$ (start date or initial time $t$ = 0) set in the simulation. The parameters $\xi_{s}$ and $\xi_{a}$ $\in$ [0,1] are measures of interaction between the susceptible and infected population (symptomatic and asymptomatic respectively);  $T$ denotes the timescale that determines how fast the effect of lockdown on infection transmission becomes prominent. $\xi_{s}$ = $\xi_{a}$ = 1 means that there is no lockdown; the infected population is freely interacting with the susceptible population and exposing the susceptible individuals to the disease at rates $\beta_{s}$ and $\beta_{a}$. $\xi_{s}$ = $\xi_{a}$ = 0 is the idealized scenario of lockdown where all the infected individuals (both symptomatic and asymptomatic) are isolated/quarantined. Hence, any sort of interaction of the infected individuals with the susceptible population is ruled out. The factors $1- \xi_{s}$ and $1- \xi_{a}$ characterize the asymptotic reduction of $\bar{\beta_{s}}(t)$ and $\bar{\beta_{a}}(t)$ due to intervention/containment measures.

The time-dependent form of $\bar{\beta_{s}}(t)$ and $\bar{\beta_{a}}(t)$ is a non-unique choice consistent with the overall trend of the disease progression and tuned reasonably to arrive at the best fit with the available data. In the current form of $\bar{\beta_{s}}(t)$ and $\bar{\beta_{a}}(t)$, when $T$ is considerably large compared to $t-\tau$ (and $\phi = 0$ ), we revert to the classic SEIRD model for symptomatic growth with a constant $S \rightarrow  E$ transition rate $\beta_{s}$. Smaller values of $T$ mark faster mitigation of $\bar{\beta_{s}}(t)$ and $\bar{\beta_{a}}(t)$. Although the choice of exponential decay is arbitrary, it has been exploited in prior COVID-19 modeling studies \cite{Fanelli2020, Chatterjee2020, Fernandez2020} and HIV/AIDS models \cite{Chowell2016, Granich2009}. We also tested other predetermined, time-dependent functional forms of $\bar{\beta_{s}}(t)$ and $\bar{\beta_{a}}(t)$ such as compressed exponential and linearly decaying functions. The model outcomes/features qualitatively remain unchanged subjected to these choices.
To determine $\bar{\beta_{s}}(t)$ (and $\bar{\beta_{a}}(t)$), a more effective choice would be to compute/extract a dynamic $\bar{\beta_{s}}(t)$ (and $\bar{\beta_{a}}(t)$) from the reported infection data  instead of regulating the time evolution via any predetermined functional form. A recent study on COVID-19 pandemic using SIRD model proposes a novel algorithm of dynamically evaluating the transmission rate from reported deaths \cite{Fernandez2020}. Adopting a similar protocol in our SEIRD model and exploring the disease progression in the Indian context would be an interesting future endeavour.

During the lockdown, $\beta_{s}$ and $\beta_{a}$ are modified as depicted in \Cref{eq9,eq11}. Hence, the mathematical form of $R_{e}$ during the lockdown period stands as
\begin{equation}
R_{e} = \frac{\phi \bar{\beta_{s}}}{\delta + \gamma_{s}} + \frac{(1-\phi) \bar{\beta_{a}}}{\gamma_{a}}    
\label{eq13}
\end{equation}
In longer time limit, $R_{e}$ becomes,
\begin{equation}
R_{e} = \frac{ \xi_{s}\phi \beta_{s}}{\gamma_{s} + \delta} + \frac{ \xi_{a} (1-\phi) \beta_{a}}{\gamma_{a}} 
\label{eq14}
\end{equation}

It is evident from \Cref{eq14} that, the first term in the expression of $R_{e}$ concerns about the disease transmission via the symptomatic carriers whereas the second term deals with the transmission through asymptomatic carriers. In a population under lockdown, it is plausible to mitigate $R_{e}$ through the gradual decrease in $\xi_{s}$ and $\xi_{a}$, the interaction parameters. In essence, this is consistent with the default definition of $R_{e}$; diminishing the interaction between the susceptible and the infected reduces $R_{e}$ and vice-versa.

Since the constituting compartments of the model are $S$, $E$, $I_{s}$, $I_{a}$, $R_{s}$, $R_{a}$ and $D$ (Fig. \ref{fig1}B-\ref{fig1}C), from now on, we can call the model adopted here, with the acronym of $SEI_{s}I_{a}R_{s}R_{a}D$ model.

All the relevant parameter values are listed in Table \ref{tableS1}-\ref{tableS2}.

\section*{Results}
\begin{figure}[htbp!]
\includegraphics[width=\linewidth]{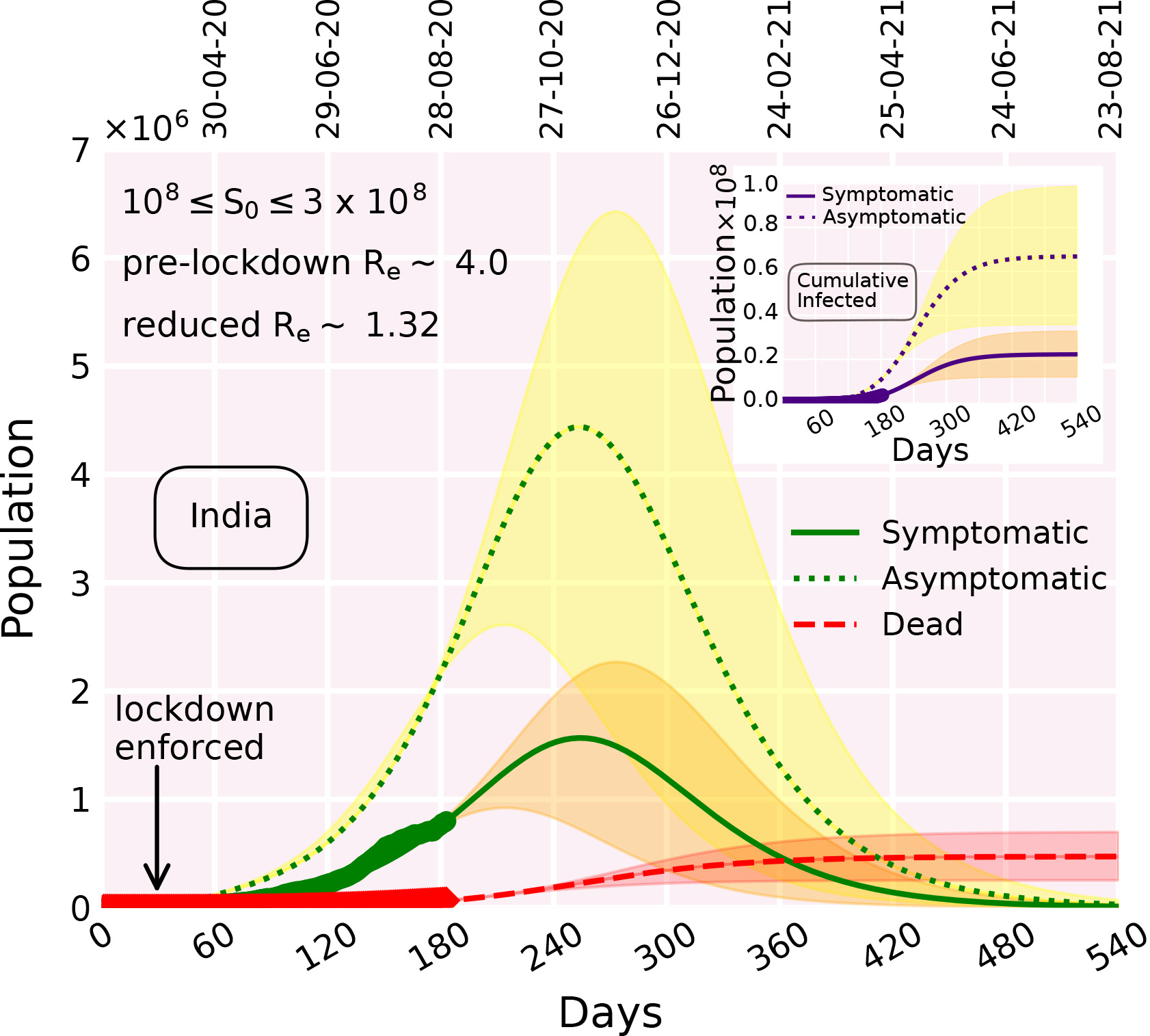}
\caption{Time evolution of the population of symptomatically infected (solid curve), asymptomatically infected (dotted curve) and dead (dashed curve) in India. The initial susceptible population is varied within a range 100-300 million. The color shades encasing the curves indicate the variation in the initial susceptible population $S_{0}$. (inset) Time evolution of the cumulative population of symptomatic and asymptomatic infections. The cumulative population of symptomatically infected is given by $I_{s} + R_{s} + D$. Similarly, cumulative asymptomatic population is given by $I_{a} + R_{a}$. The real data (plotted with points) are considered from March 2, 2020 up to September 1, 2020.} \label{fig2}
\end{figure}

We investigated the role of asymptomatic population propagating COVID-19 infection in India and various states within the country using the $SEI_{s}I_{a}R_{s}R_{a}D$ model. The results are elucidated in the following and represented in Fig. \ref{fig2}-\ref{fig6}. To begin with, we varied the initial susceptible population $S_{0}$ within a range of 100-300 million for India (Fig. \ref{fig2}) and explored how the infected population is getting subdivided into two categories: (a) symptomatic population and (b) asymptomatic population; moreover how the asymptomatic population is contributing to the recent surge of infections in India (Fig. \ref{fig1}, Fig. \ref{fig2}-\ref{fig3}). Next, we explored how the symptomatic and asymptomatic infection peaks are correlated and whether detecting and quarantining the symptomatic population is sufficient enough to contain the disease spread or the `hidden' carriers without any symptoms make the disease transmission difficult to contain (Fig. \ref{fig3}A-\ref{fig3}B). Furthermore, we briefly demonstrate how the effective reproduction number $R_{e}$ computed from the current model behaves under the effect of lockdown/containment measures (Fig. \ref{fig4}). Next, we explore the symptomatic and asymptomatic infection growth curves for a few Indian states using the $SEI_{s}I_{a}R_{s}R_{a}D$ model (Fig. \ref{fig5}). Lastly, we conclude by examining the COVID-19 data for cumulative infections and deaths in Indian context in the light of Benford's law \cite{Benford1938, Newcomb1881, Hill1998, Sambridge2020} projections (Fig. \ref{fig6}).

\subsection*{Asymptomatic population and the recent surge of infections: a worrisome  affair for India }
We first investigated how the infection growth curves would look like in India's context if we consider asymptomatic infections. We start with an effective reproduction number $R_{e}$ $\sim$ 4 at the onset of the simulation (Fig. \ref{fig2}). As the lockdown is enforced during the initial stages of the outbreak, $R_{e}$ reduces to value $\sim$ 1.32. The initial susceptible population is varied within a range of 100-300 million. During the lockdown period, the interaction parameters are chosen as $\xi_{s}$ $\sim$ 0.27 and $\xi_{a}$ $\sim$ 0.36 for best fit. $\xi_{s}$ $\sim$ 0.27 and $\xi_{a}$ $\sim$ 0.36 imply that the interaction between the susceptible and the symptomatically infected population is reduced to $\sim$ 27 $\%$ during the lockdown; similarly, the interaction between the susceptible and the asymptomatically infected population is reduced to $\sim$ 36 $\%$. The model analysis shows that the total number of asymptomatic infections is several-fold higher than the symptomatic infections (which is of course subjected to the choice of fitting parameters). In the following section, we discuss the relative fraction of symptomatic and asymptomatic population from the model perspective.

As expected, with an increase in the initial susceptible population $S_{0}$, the infection peaks also rise to greater heights. The symptomatic and asymptomatic infection growth curves appear to reach the respective peaks in the middle of November 2020 (Fig. \ref{fig2}).  With the increase in $S_{0}$, the peaks shift to a later time point (Fig. \ref{fig2}). Note that, both the infection peaks may occur at the same time point as depicted in Fig. \ref{fig2}. The cumulative infection curves (inset, Fig. \ref{fig2}) show a flattening signature from the end of December.

The question that intrigues next, is how robustly do the model predictions zero in on the relative numbers of symptomatic and asymptomatic infections?

\subsubsection*{Sensitivity of the model projections on the relative sizes of symptomatic and asymptomatic populations}

From various public health bulletins and media reports, we gather that number of asymptomatic infections is possibly much higher than the symptomatic. Since the asymptomatic individuals do not develop symptoms (or have mild symptoms), they are harder to detect. In the Indian context, even though the testing capacity has been significantly increased during the past few months, the testing strategy is largely concentrated on individuals showing typical symptoms. To that effect, attempting to find asymptomatic cases by contact tracing and random testing raises the possibility that many of the asymptomatic carriers will go undetected. Hence, there exists a significant paucity of data about the exact number of asymptomatic cases compared to the symptomatic ones. This uncertainty leaves us with a finite window of variability in the parameter space concerning the model analysis.

\begin{figure}[htbp!]
\includegraphics[width=0.75\linewidth]{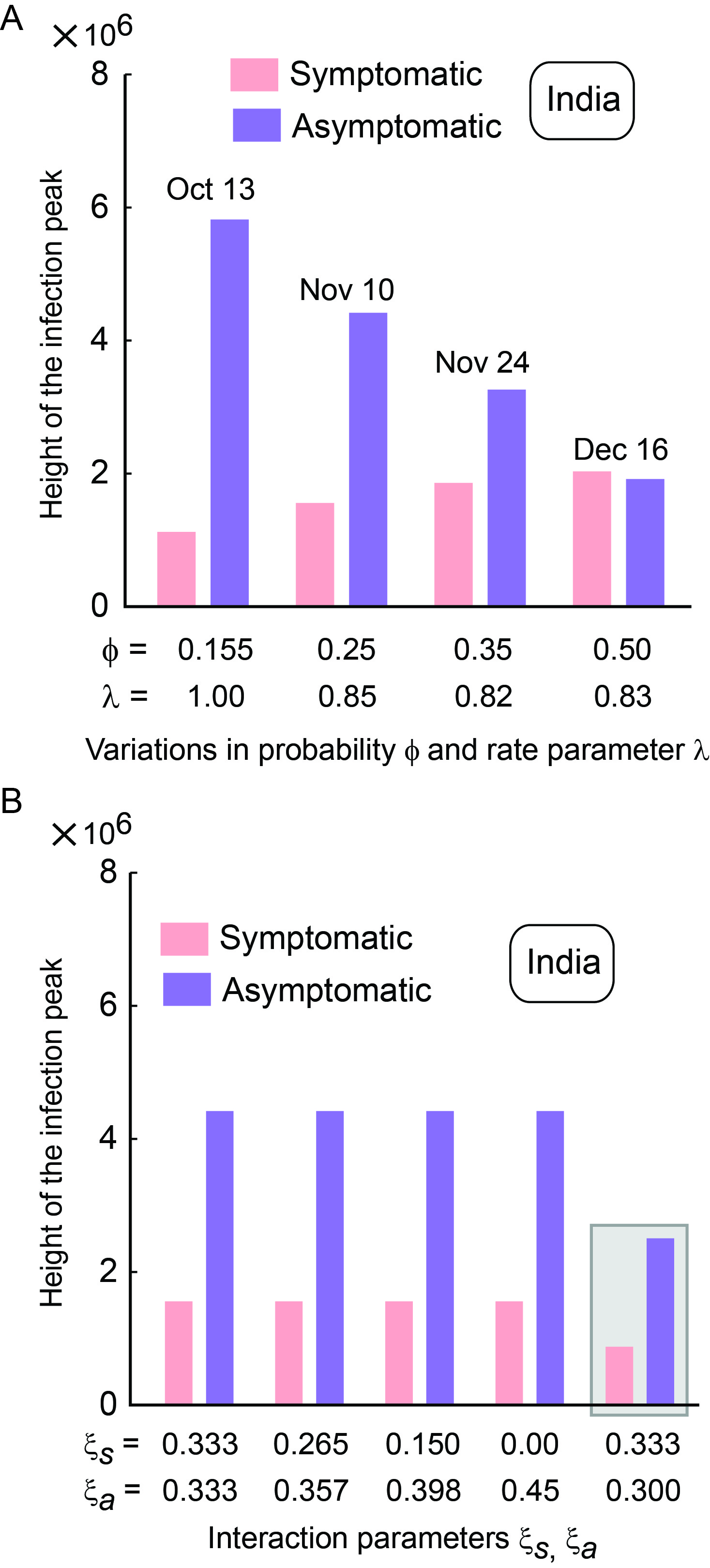}
\caption{Additional parameter dependence of symptomatic and asymptomatic infection peaks. (A) Symptomatic and asymptomatic infection peak heights (and tentative time window around which the peaks occur) depend upon the choices of probability $\phi$ and rate parameter $\lambda$. (B) Non-unique choices of symptomatic and asymptomatic interaction parameters $\xi_{s}$ and $\xi_a$ may yield identical infection curves (not shown) and identical peak heights at the same time. The peak heights of the infection curves are obtained from the best fit with the real data for different values of $\xi_{s}$ and $\xi_{a}$, keeping other parameters fixed at base values. 
The last set of bars shaded in grey indicates that the infection peaks would be much lower than the projected trend if the asymptomatic patients are detected and quarantined (manifested through the lower value of $\xi_{a}$). For all cases, the initial susceptible population $S_{0}$ is chosen to be about 200 million. The real data for fitting is considered from March 2.} \label{fig3}
\end{figure} 

If we inspect \Cref{eq1,eq2,eq3,eq4} and Fig. \ref{fig1}B closely, we find that the exposed population is `fed' into the compartments of the symptomatic and asymptomatic population with probability rates $\phi \lambda$ and $(1-\phi)\lambda$ respectively. The choices for the numerical value of $\phi$ ($\phi$ $\in$ [0,1]) regulates the relative size of the symptomatic and asymptomatic populations (since $\gamma_{s} \sim \gamma_{a}$, the recovery time for symptomatic and asymptomatic populations are similar). If $\phi \sim 0.5$, the peaks for the symptomatic and asymptomatic infections are likely to have the same height, as per the default model construction (Fig. \ref{fig3}A). The fact that the number of asymptomatic infections likely to be several folds higher than the symptomatic infections \cite{Kissler2020} prompted us to make the current choice of the numerical value of $\phi$. The greater size of the asymptomatic population, as suggested in various reports, alludes that it is reasonable to restrict $\phi$ in the range 0.2-0.4. The best fit with real data shows that, as $\phi$ increases, the peak of the symptomatic infections consistently climbs up to a certain value determined by the current trend (Fig. \ref{fig3}A). In Indian context, we have chosen $\phi = 0.25$, unless otherwise mentioned (Table \ref{tableS1}, \ref{tableS2}). Note that in the default model construction, what fraction of the net outflux from the `exposed' compartment ($E$), would go into the symptomatic ($I_{s}$) and asymptomatic ($I_{a}$) compartments depends on the factors $\phi$ and $1-\phi$ respectively. Therefore, when $\phi = 0.25$, we expect that the peak of the asymptomatic growth would be $\sim$ 3-fold higher than that of symptomatic growth. This is evident from Fig. \ref{fig2}, \ref{fig3}A. 

The next question is, whether one can attribute the recent surge in the number of symptomatic infections to the largely undetected asymptomatic population. 

\subsubsection*{The recent surge in symptomatic infections: Does the onus fall on the asymptomatic carriers?}
To address this question from the current model perspective, we visit the \Cref{eq1,eq2} and Fig. \ref{fig1}B-\ref{fig1}C. Imagine that all the active cases with symptoms are detected and quarantined. Then, the interaction among the susceptible and the symptomatic population becomes almost null as $\bar{\beta_{s}}$ effectively falls close to zero. But, even in that scenario, the `exposed' ($E$) compartment can still be `fed' through the interaction between susceptible and the asymptomatic individuals at a rate $\bar{\beta_{a}}$. This exposed population, in turn, fluxes into symptomatic and asymptomatic compartments with probability rates $\phi \lambda$ and $(1-\phi)\lambda$. Thus, it is clear that merely quarantining the symptomatic solely cannot prevent the surge in both symptomatic and asymptomatic infections. The mixing of asymptomatic individuals with the susceptible may lead to new infections which are both symptomatic and asymptomatic.
In the Indian context, to date, the testing capacity is mostly dedicated to the detection of symptomatic candidates. Once a positive case with symptoms is detected, the infected person is isolated and quarantined along with others who are traced to come in contact with the infected person. But the cases that show no symptoms are largely undetected. In most of the asymptomatic cases, the person her/himself is unaware of the fact that s/he is transmitting the disease. Thus, even during the lockdown, the mixing of an asymptomatic person with the susceptible cannot be fully filtered and prevented. From the model perspective, this may be reflected in the numerical values of symptomatic and asymptomatic interaction parameters $\xi_{s}$ and $\xi_{a}$, during the lockdown. Extensive detection and quarantining of the cases with symptoms may reduce $\xi_{s}$ significantly. However, since the asymptomatic population remains largely untapped, $\xi_{a}$ may not decrease that much compared to $\xi_{s}$ during the lockdown. In brief, the infection curves will continue to surge if only the symptomatic population is put into quarantine (reflected in significantly suppressed $\xi_{s}$) but asymptomatic individuals roam freely (gleaned from the relatively marginal reduction in $\xi_{a}$) during the lockdown. Interestingly, we notice that non-unique choices of $\xi_{s}$ and $\xi_{a}$ (while other relevant parameters are kept fixed at base values) may lead to identical best fits with the real data. If $\xi_{s}$ is decreased and simultaneously $\xi_{a}$ is increased, the combinatorial effect of these two interaction parameters results in identical best-fit curves with precisely the same infection peak height and peak location (Fig. \ref{fig3}B). From a physical perspective, we can argue that even if all the symptomatic carriers are detected and quarantined, it is not sufficient to suppress the infection growth. A reasonably higher value of $\xi_{a}$ (asymptomatic carriers mixing with susceptible) can compensate for the isolated symptomatic population (lower $\xi_{s}$, symptomatic infections barred to mix with susceptible) and single-handedly keep on fuelling the infection growth at a significant rate (see the first four set of bars, Fig. \ref{fig3}B).
It is also evident that if $\xi_{a}$ is significantly reduced by extensive and/or randomized testing and quarantining patients without symptoms, the infections may be contained at a much lower number (see the last set of bars shaded in grey, Fig. \ref{fig3}B).

During the initial stages of the pandemic (April-May, 2020), once the lockdown was enforced, India had observed a surge of reverse migration of migrant workers from one part of the country to another. Preliminary thermal screening at the destination of their journey, e.g at rail/bus stations, could diagnose symptomatic patients only. However, a large number of these people were possibly asymptomatically infected. Due to a lack of social distancing during their plighted journey, the chance of infection propagation increased manifold. The rapid increase of infections visible in Fig. \ref{fig1}A (bars enclosed in dashed box) might be linked with the reverse migration. Several recent studies highlight the effect of human mobility on the COVID-19 progression in detail \cite{Kraemer2020, Galeazzi2020, Biswas2020a}.       

\begin{figure}[htbp!]
\includegraphics[width=\linewidth]{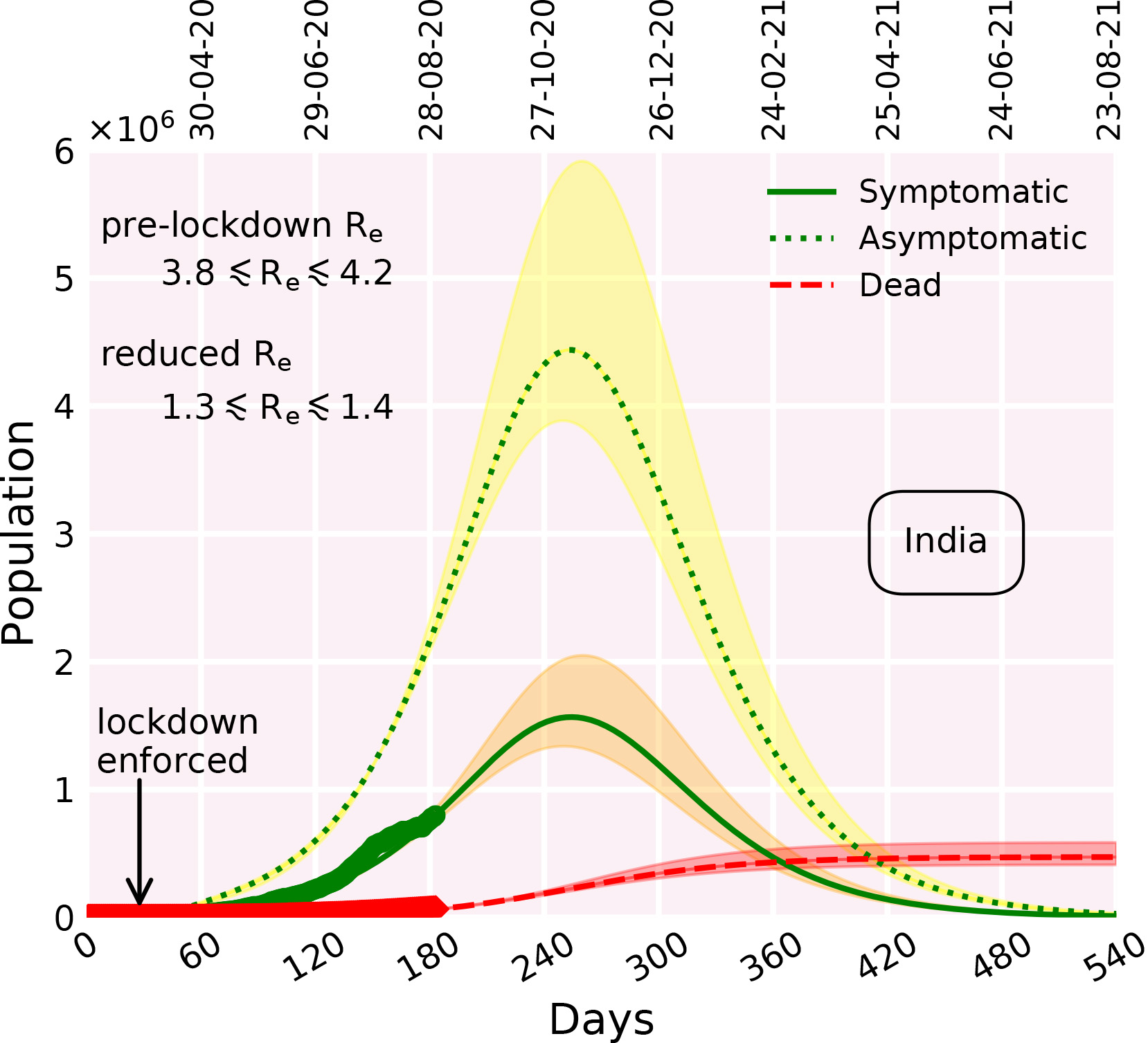}
\caption{Best fit of the symptomatic infection and death curves along with the projected asymptomatic infection curve when the pre-lockdown value of the effective reproduction number $R_{e}$ is tuned within a range 3.8-4.2. The color shades encasing the curves indicate the variation in the effective reproduction number $R_{e}$. The initial susceptible population is chosen to be about 200 million. The real data (represented as points) are plotted from March 2.} 
\label{fig4}
\end{figure}

Next, we discuss how the effective reproduction number $R_{e}$ computed from the current $SEI_{s}I_{a}R_{s}R_{a}D$ model evolves during the lockdown. 

\subsubsection*{How does the lockdown affect the effective reproduction number $R_{e}$?}

During the lockdown, the interaction parameters $\xi_{s}$ and $\xi_{a}$ are suppressed because the enforcement of containment measures and social distancing reduces the mixing of infected and susceptible. It is evident from \Cref{eq14} that, as $\xi_{s}$ and $\xi_{a}$ decrease, the pre-lockdown $R_{e}$ should also plummet down to a lower value indicating a relatively slower speed of infection.

In Fig. \ref{fig4}, we observe that the pre-lockdown $R_{e}$ decreases significantly during the lockdown. Also, in consonance with the definition of $R_{e}$ \cite{Chatterjee2020}, we find that the higher the value of $R_{e}$, the greater the number of infections as gleaned from the peak heights (Fig. \ref{fig4}). However, in order to convincingly flatten the infection curve, $R_{e}$ needs to be taken down to a value below 1. This is a challenge that we have to tackle with utmost priority and sincere policy-making.  

Next, we explore the COVID-19 progression in a few Indian states and the influence of the `hidden' asymptomatic population on the infection growth using this $SEI_{s}I_{a}R_{s}R_{a}D$ model. 

\begin{figure*}[htbp!]
\includegraphics[width=0.85\linewidth]{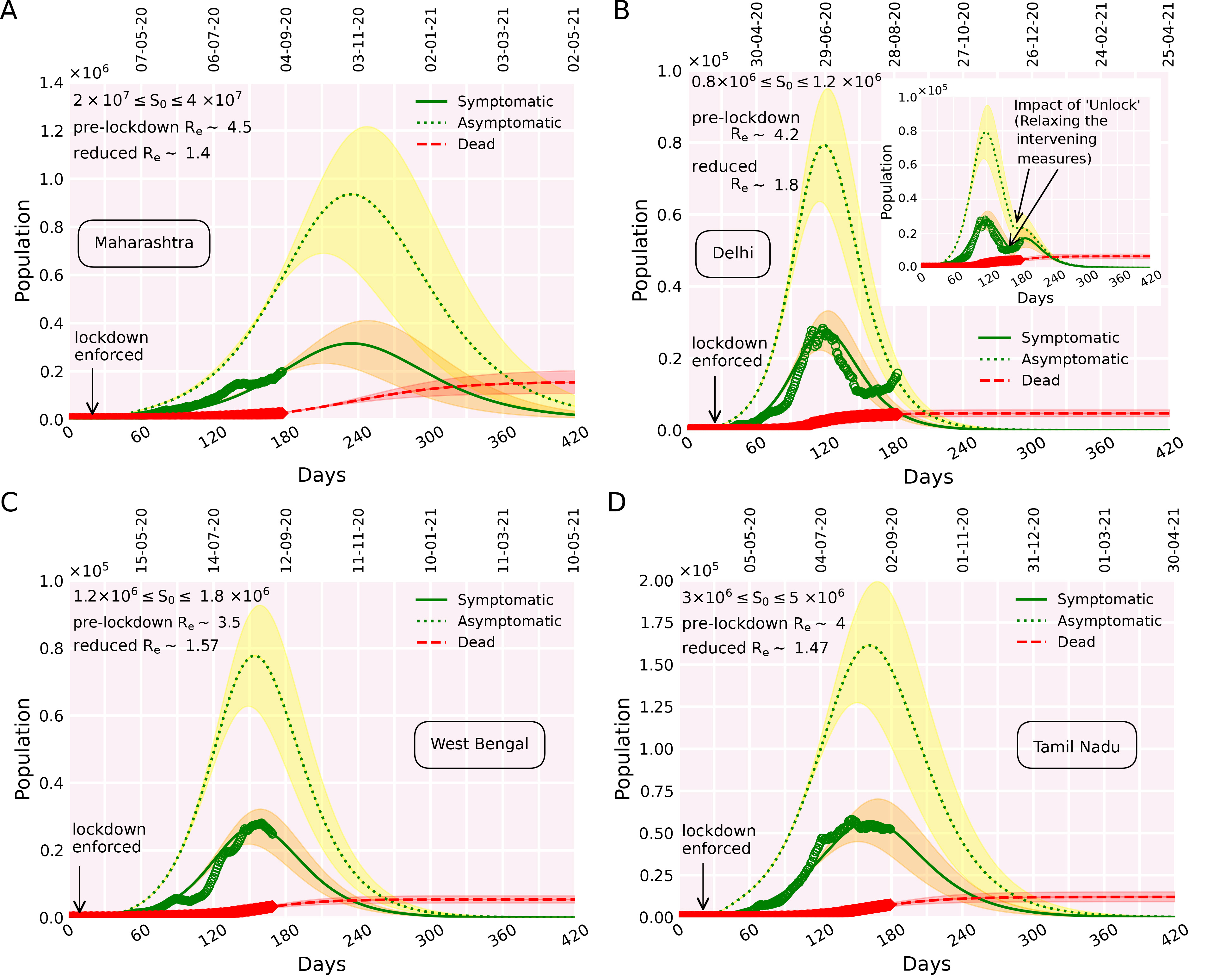}
\caption{Best fit of the symptomatic infection and death curves along with the projected asymptomatic infection curves for Indian states Maharashtra (A), Delhi (B), West Bengal (C) and Tamil Nadu (D).} \label{fig5}
\end{figure*} 

\subsubsection*{Asymptomatic carriers and COVID-19 propagation in few Indian States}
What we observe from the best fit of the infection curves for Indian states Maharashtra, Delhi, West Bengal, and Tamil Nadu is that in most of the states (except Maharashtra), the infections curves have already attained the respective peaks \ref{fig5}A-\ref{fig5}D. For Maharashtra, the active infection continues to surge. The infection peak is expected towards the end of October/beginning of November 2020, if the current trend (mapped from the real data) continues (Fig. \ref{fig5}A). In the case of Delhi, towards the end of July 2020, the number of active infections was significantly reduced due to strict intervening measures (Fig. \ref{fig5}B). However, gradual `unlock' and resuming the social and economic activities, interventions effectively relaxed over time. To that effect, the active infections have started to increase again, lately (Fig. \ref{fig5}B). To account for the resurgence as visible from the real data into the theoretical infection curve, we resorted to the following: (a) after the lockdown is enforced, $\bar{\beta}_{s}$ and $\bar{\beta}_{a}$ follow the exponential decline as noted in \cref{eq9,eq11}; (b) $\bar{\beta}_{s}$ and $\bar{\beta}_{a}$ are governed by \cref{eq9,eq11}, till $\sim$ 1 month after the peak infection is attained. The $\sim$ 1-month time window is set by the fact that after attaining the peak, the number of active infections was still on the decline. However, 5-6 weeks after the peak infection, active infections started to rise again gradually. To match the trend evident from the real data, we relaxed $\bar{\beta}_{s}$ and $\bar{\beta}_{a}$ in the model, mimicking the effective lifting of intervening measures during the `Unlock' phases. Once the active infections began to re-surge, $\bar{\beta}_{s}$ and $\bar{\beta}_{a}$ were exponentially increased as opposed to the pattern set by \cref{eq9,eq11}. Upon these considerations, we observe that the theoretical curve is in reasonable agreement with the available real data (Fig. \ref{fig5}B, inset).

Notably, another consistent feature is that the predominantly `hidden' asymptomatic infections may be several folds higher than the symptomatic infections in all the cases. This, of course, in the model, depends on the fraction determined by $\phi$ as discussed previously (Fig. \ref{fig3}A, \ref{fig1}B-\ref{fig1}C).

\section*{Summary and discussion}
The spread of COVID-19 via presymptomatic and asymptomatic cases has been a big concern in recent times as the mobility of people is increasing while lockdown is being relaxed across the country. Since presymptomatic and asymptomatic persons, having no influenza-like symptoms, are not aware of their potential of infecting others, a relaxing lockdown gives ample opportunity to expose a contained population to the virus that effectively increases the number of susceptible. Therefore, the projected trajectories of infections (particularly the infection peaks) largely depend upon the number of susceptible ($S_{0}$) in the model. The choice of the initial size of the susceptible population  $S_{0}$ is a tricky business. Theoretically, in the absence of immunity (acquired or otherwise), an entire population may become susceptible to the COVID-19 outspread. However, in reality, all people will rarely be susceptible to the disease no matter how rapidly it spreads. Various factors like geography, socio-economic characteristics, the demographic landscape can significantly regulate $S_{0}$. The estimation of a `ball-park number' for $S_{0}$ is crucial to initiate the modeling analysis. To estimate a plausible value of $S_{0}$, first, we look at the infection curves for the countries where the cumulative positive cases have already moved past the peak infection and attained a plateau. Dividing the number of infections at the plateau by the total population of the country under consideration yields a fraction that reflects the average percentage of the population infected by the virus. This fraction turns out to be $\sim$ $10^{-3}$ for large countries like Germany, USA, Spain, Italy \cite{Chatterjee2020}. Thus, an estimate of susceptible can be computed by multiplying this fraction with the total population of a country. For India, the total population is in the order of $\sim 10^{9}$. Hence, a rough `ball-park' number for the initial susceptible population $S_{0}$ would be in the range of $10^{9} \times 10^{-3}$ $\sim$ $10^{6}$, i.e in the order of millions. This estimation for $S_{0}$ gives us a lower bound. The reason is, in a hugely populated and large country like India, even if the testing capacity is increased, due to limited resources the testing strategy probes individuals having symptoms in the first place. Thus, many mildly symptomatic and asymptomatic cases remain undetected. Therefore, the numbers gauged from the cumulative infection plateau allude to an underestimation up to a certain degree. 
An alternative approach finding an approximate $S_{0}$ would be to compute the ratio between the total number of positive cases (i.e the number of cumulative infections) and overall tests carried out. The ratio essentially is the test positivity rate. Multiplying this ratio with the total population of a country gives another `ball-park' estimation of $S_{0}$ or approximately the order of magnitude for $S_{0}$ \cite{Chatterjee2020}. For example, in the case of India, the estimation from test positivity rate yields an order of magnitude $\sim 10^{8}$ for $S_{0}$. Normally, in these kinds of compartmentalized epidemiological models, $S_{0} \sim N$. To attain the best fit with the reported data, we varied $S_{0}$ within a range gleaned from the above estimations in a context-dependent manner for India and a few Indian states.

In the context of COVID-19 infection propagation, a few interesting aspects to ponder over are: (a) whether there is any bias in people getting infected from asymptomatic patients to be asymptomatic and (b) whether the rates of infection from symptomatic and asymptomatic patients are the same. About (a), it may seem intuitive that a person getting infected from an asymptomatic would turn out to be asymptomatic as well. But, in reality, the immune response of an infected individual can be very different. Presently, to our knowledge, there is no conclusive data to assess this point. About (b), we discuss a few key features of infection propagation through asymptomatic and symptomatic carriers in the following. An asymptomatic individual is likely to carry less viral loads compared to a symptomatic individual. Since the droplet transmission (via coughing, sneezing) is a dominant mode of infection propagation, the chance of infection via an asymptomatic individual is less likely than a symptomatically infected person since an asymptomatic carrier hardly develops conspicuous symptoms like coughing, sneezing, etc.. On the contrary, asymptomatic individuals lead `normal' social life of human interactions without being aware that they are `silent' carriers whereas symptomatic individuals are more likely to be under restriction mixing with people. Therefore, within a population, chances of disease spread via asymptomatic carriers may be higher.

In the model analysis, for simplicity, we have chosen the values of $S \rightarrow E$ transition rates $\bar{\beta}_{s}$ and $\bar{\beta}_{a}$ to be the same (Table \ref{tableS1}-\ref{tableS2}). But what fraction of $E$ emerges out as symptomatic or asymptomatic is regulated by $\phi$. Note that, in the model construction, there is no bias in people getting infected from a/symptomatic patients to be a/symptomatic. In the `exposed' compartment $E$, the population is a mixed due to (a) $S-I_{s}$ interaction and (b) $S-I_{a}$ interaction. But no information about the mode of infection via (a) and (b) are book-kept in the total exposed population $E$. Therefore, individually, neither (a) nor (b) can influence the division in the outflux from $E$ into $I_{s}$ and $I_{a}$ (Fig. \ref{fig1}B, \cref{eq1,eq2,eq3,eq4}). 

We have varied $S_{0}$ for India within a range 100-300 million (Table \ref{tableS1}, Fig. \ref{fig1}B). Note here, that this estimation is a simplistic approximation. After all, it is difficult to comment on the percentage of the population infected by the virus until the pandemic is over. Further details regarding the estimation of $S_{0}$ is described in our earlier study (\cite{Chatterjee2020}).

In brief, as per the current trend of the symptomatic cases, India is expected to see a peak in the active infection in November 2020 (Fig. \ref{fig2}), tentatively. The number of active infections at the peak may lie somewhere between 1.5- 2.5 million.  The total infection (cumulative) would then reach around 10 – 20 million (mid-November) as shown in Fig. \ref{fig2}, inset. Note that the COVID-19 pandemic is an ongoing phenomenon with its time evolution depending upon various local and global factors. Therefore, quantitative predictions from a simplistic model investigated in this article, is expected to change as the pandemic progresses.

\begin{figure}[htbp!]
\includegraphics[width=\linewidth]{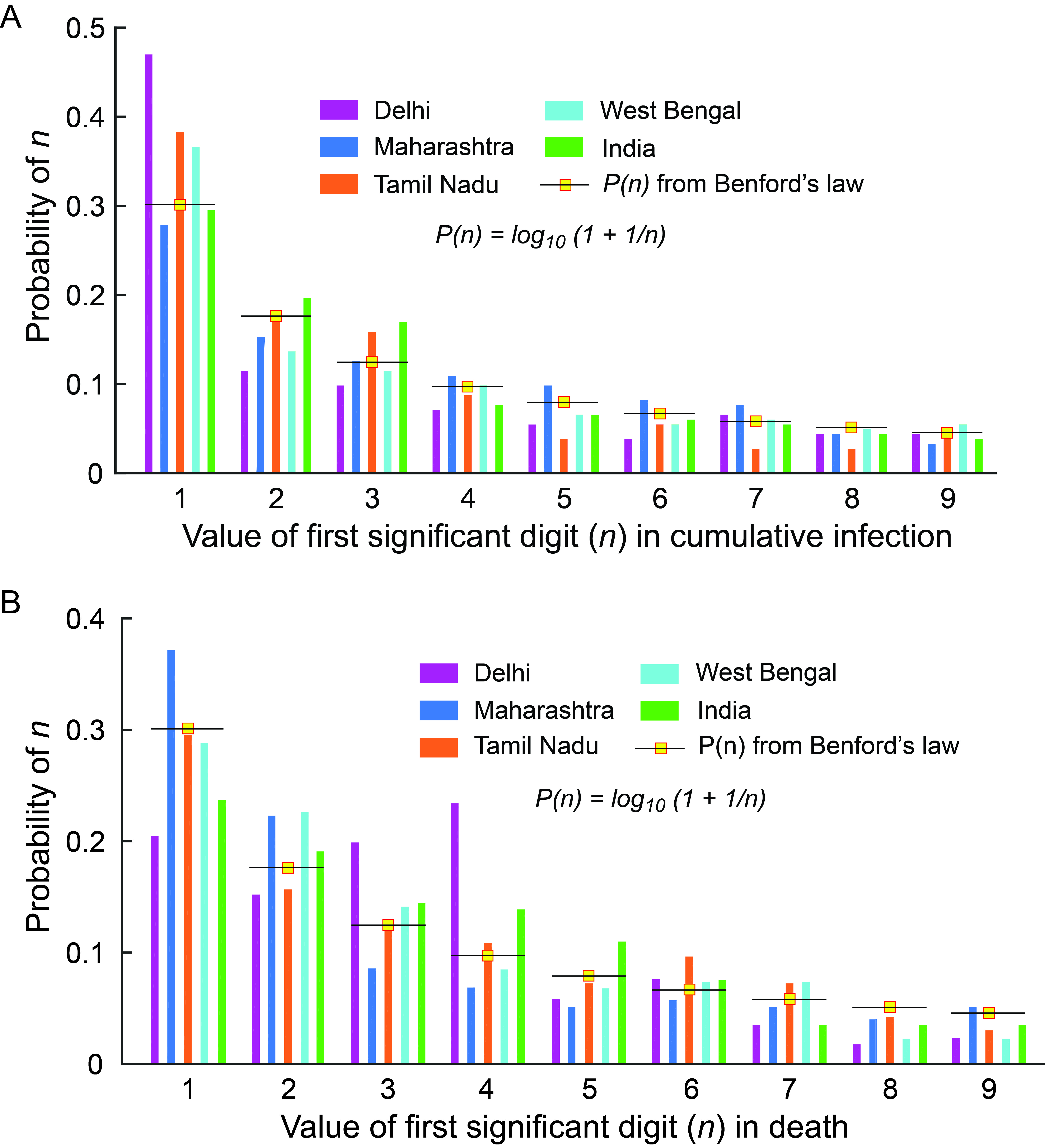}
\caption{Data for cumulative infection and death due to COVID-19 pandemic in few Indian states and India as a whole in the light of Benford's law (BL) projections. (A-B) The probability distributions of first significant digits $n$ ($n$ $\in$ [1, 2,...,9]) appearing in the data chart of cumulative infection (A) and death (B) are plotted as histograms. The probability of $n$, according to BL is given by $P(n) = log_{10} (1 + \frac{1}{n})$. The overall trend as observed from the reported data for India reasonably follows the projections from BL.
} \label{fig6}
\end{figure} 

The quality of the COVID-19 data (e.g numbers representing cumulative infection, death, etc.) for various countries, provinces and local regions made available in the public domain is a crucial determinant for the study of the pandemic progression. A frequently used tool for testing the reliability of an empirical data set is Benford's law (also known as Newcomb-Benford law) \cite{Benford1938, Newcomb1881, Hill1995, Hill1998}. The phenomenological Benford's law (BL) states that in diverse sets of `real world'/`naturally occurring' numerical data, the first significant digit $n$ appears with probability $P(n) = log_{10} (1 + \frac{1}{n})$  \cite{Benford1938, Newcomb1881, Hill1995, Hill1998}. In this study, we compared the data of cumulative infections and deaths for India and few Indian provinces (e.g Delhi, Maharashtra, Tamil Nadu, West Bengal) in the light of BL and tested the compliance (Fig. \ref{fig6}A-\ref{fig6}B). It is important to note that while many numerical data sets comply with BL, several collections of numbers do not. Therefore, only checking the compliance (or deviation) of a distribution of the first significant digit with  $P(n)$ is not a sufficient condition to ascertain the reliability of a data set. Furthermore, it is also discussed that in case of a certain number of individual distributions that do not have significant digit frequencies close to those determined by BL, random sampling from those distributions may result in a combined collection of numbers with significant digit frequencies obeying BL \cite{Hill1998}. Recent studies on the quality of COVID-19 data suggest that better compliance with BL occurs when the cases are steadily on the rise during the infection progression in a region \cite{Sambridge2020, BokLee2020}. At later stages, when the implementation of containment/intervening measures flattens the infection curves, the data sets are unlikely to follow BL.

In this current study, we have included the gradual relaxation of lockdown measures (`unlock 1.0' plan starting from June 1, laid out by Govt. of India and at present it is unlock 4.0), only in the case of Delhi, where the active infection data shows a resurgence after the attainment of the peak. In the Indian context as a whole, as the `Unlock' phases are initiated, we expect a considerable surge in the active cases in the upcoming days. We assessed the COVID-19 progression in the presence of `one-time enforcement of containment measures' where the lockdown and other social distancing norms remain in place for an indefinite time. However, a more realistic reconstruction of the pandemic situation would be to impose `intermittent lockdown' in specific regions where the containment and other social distancing measures are enforced once the number of active infections crosses a threshold determined by the capacity of the regional healthcare system. Afterward, the lockdown measures are relaxed/lifted as the active infections fall below a certain threshold (`unlocking' the lockdown). This shuttling between the phases of `lockdown' and `unlock' continues until the contagion comes totally under `control' or the threat of an `out-of-bound' infection is eliminated. The nature of intermittent intervention depends on various controlling factors like acquiring `herd immunity', availability of proper vaccines, the capacity of public health facilities where all the patients can be accommodated and treated, etc. A previous study (\cite{Kissler2020}) in the context of the USA, has discussed these aspects. In the Indian context, it would be interesting to explore further along this avenue as a future venture.       

Like the SIRD model used in our previous study \cite{Chatterjee2020}, this $SEI_{s}I_{a}R_{s}R_{a}D$ model does not accommodate any spatial information about the infection spread. In other words, there is no spatial degree of freedom in the governing equations of $SEI_{s}I_{a}R_{s}R_{a}D$ model. The spatial dependence of the disease progression can be executed via simulating this model on a lattice or connected network. The network may feature the following attributes: (a) every node on the network virtually carries the territorial/regional information collected from the map of India; (b) the network connectivity bears the details of human mobility, transmission spread from one region to another. The network nodes also can account for the topology/geography of India in a coarse-grained manner. For example, the nodes representing Himalayan highlands would likely to be attributed to lesser human mobility and transmission spread compared to the nodes representing Indian metro cities. This proposed model would enable us to study how the infection spreads from a few initial local pockets to a larger region, thereby rendering a broader picture of the dynamics of the pandemic. A previous study investigating the COVID-19 progression in Italy describes a novel approach using a spatially explicit SEIR model \cite{Gatto2020}. Future investigation with similar spatial/topological detailing in India's context would be useful to delve into. Also, note that the current $SEI_{s}I_{a}R_{s}R_{a}D$ model is a generalized model. It does not contain any specific biological/clinical features of the COVID-19 disease. In this model, COVID-19 enters through the rates and governing parameters in \Cref{eq1,eq2,eq3,eq4,eq5,eq6,eq7,eq8,eq9,eq10,eq11,eq12} that are tuned to obtain the best fit of the theoretical curves with the available COVID-19 data.

In the currently explored model parameter regime, we observed that the number of asymptomatic cases is several-fold higher than the symptomatic cases. In reality, even though the number of tests has been increased significantly, individuals showing symptoms are given priority to undergo tests. In a hugely populated country like India, this leaves a large fraction of the asymptomatic population untraceable. Therefore, to determine the exact ratio of symptomatic and asymptomatic population, in reality, we need extensive data curation across the country and further research in that direction. The infectiousness of an asymptomatic individual may be similar to a person having symptoms. The asymptomatically infectious population, mixing freely with the healthy susceptible population, keeps spreading the infection causing a surge in the number of symptomatic as well as asymptomatic cases. To reduce the infection, it is important to trace the source of infection and isolate it from the healthy susceptible population. Currently, there is merely any data available to validate our prediction of the asymptomatically infected population. To assess the community spreading and prevalence of asymptomatic cases, India conducts a serology-based (antibody test) survey in select districts. The test aims to find the presence of a specific antibody developed by the immune system of the infected person in response to the viral infection. If the data from the survey is made available in the public domain, the model prediction can be assessed. Irrespective of the testing policy, maintaining social distancing in tandem with prolonged or intermittent containment measures would be crucial. Besides, it is also necessary to use cloth face coverings or mask across the population.

To conclude, `indefinite lockdown' is not a solution to put an end to the COVID-19 outspread. The broader purpose, the lockdown in India served, is the opportunity to buy `time'. In a nutshell, the lockdown slows down the infection and the lockdown time window provides us the chance to ramp up the health care facilities, testing capacity, etc. so that when the lockdown is lifted, the COVID-19 does not catch us off-guard. During the gradual `Unlock' phases, the effect of lockdown can still prevail if social distancing, sanitation protocols, and proper mask-wearing in public places are adopted by every individual as the `new normal' of lifestyle.

\section*{Acknowledgments}
Sa.C. and A.S. were supported by a fellowship from the University Grants Commission (UGC), India. M.K. was supported by a fellowship from CSIR, India. Sw.C. thanks the Indian Association for the Cultivation of Science, Kolkata for financial support. R.P. thanks IACS for support and Grant No. EMR/2017/001346 of SERB, DST, India for the computational facility.



%



\beginsupplement

\begin{table*}[htbp!]
\begin{adjustwidth}{-0.0cm}{-0cm}
\centering
\begin{tabular}{ |c|c|c|c| } 
\hline
Abbreviation & Meaning & Value\\ 
\hline
& Parameters for \textbf{India} &\\
\hline
$\beta_{s}$& \makecell{rate at which susceptible becomes exposed\\upon interaction with symptomatic population} & 0.352 day$^{-1}$\\ 
\hline
$\beta_{a}$& \makecell{rate at which susceptible ($S$) becomes exposed ($E$)\\upon interaction with asymptomatic population} & 0.352 day$^{-1}$\\ 
\hline
$\lambda$& \makecell{rate at which exposed ($E$) becomes \\infectious (both symptomatic and asymptomatic)} & 0.85 day$^{-1}$\\ 
\hline
$\phi$& \makecell{probability of being symptomatically infected} & 0.25\\ 
\hline
$\gamma_{s}$& rate of recovery for symptomatic individuals & 0.083 day$^{-1}$\\
\hline
$\gamma_{a}$& rate of recovery for asymptomatic individuals & 0.09 day$^{-1}$\\
\hline
$\delta$& rate of death & 0.0018 day$^{-1}$\\ 
\hline
$\xi_{s}$& \makecell{interaction parameter for symptomatic\\ population mixing with susceptible during lockdown} & 0.265\\
\hline
$\xi_{a}$& \makecell{interaction parameter for asymptomatic\\ population mixing with susceptible during lockdown} & 0.357\\
\hline
$\tau$& \makecell{the day from which lockdown begins (counted from day 0,\\start date of simulation or initial time $t$=0)} & 36 days\\
\hline
$T$& \makecell{delay in number of days before the effect of containment\\measures on infection propagation becomes visible} & 10 days\\ 
\hline
$S_{0}$& \makecell{initial susceptible population} & 1-3 $\times$ 10$^{8}$\\
\hline
\end{tabular}
\caption{List of parameters chosen for the best fit with the real data in Indian context.}
\label{tableS1}
\end{adjustwidth}
\end{table*}

\begin{table*}[htbp!]
\begin{adjustwidth}{-0.0cm}{-0cm}
\centering
\begin{tabular}{ |c|c|c|c| } 
\hline
Abbreviation & Meaning & Value\\ 
\hline
& Parameters for \textbf{few Indian states} &\\
\hline
$\beta_{s}$& \makecell{rate at which susceptible becomes exposed\\upon interaction with symptomatic population} & 0.31-0.37 day$^{-1}$\\ 
\hline
$\beta_{a}$& \makecell{rate at which susceptible ($S$) becomes exposed ($E$)\\upon interaction with asymptomatic population} & 0.31-0.37 day$^{-1}$\\ 
\hline
$\lambda$& \makecell{rate at which exposed ($E$) becomes \\infectious (both symptomatic and asymptomatic)} & 0.74-85 day$^{-1}$\\ 
\hline
$\phi$& \makecell{probability of being symptomatically infected} & 0.25\\ 
\hline
$\gamma_{s}$& rate of recovery for symptomatic individuals & 0.07-0.085 day$^{-1}$\\
\hline
$\gamma_{a}$& rate of recovery for asymptomatic individuals & 0.074-0.091 day$^{-1}$\\
\hline
$\delta$& rate of death & 0.0018-0.003 day$^{-1}$\\ 
\hline
$\xi_{s}$& \makecell{interaction parameter for symptomatic\\ population mixing with susceptible during lockdown} & 0.25-0.382\\
\hline
$\xi_{a}$& \makecell{interaction parameter for asymptomatic\\ population mixing with susceptible during lockdown} & 0.3325-0.47\\
\hline
$\tau$& \makecell{the day from which lockdown begins (counted from day 0,\\start date of simulation or initial time $t$=0)} & 17-30 days\\
\hline
$T$& \makecell{delay in number of days before the effect of containment\\measures on infection propagation becomes visible} & 5-18 days\\ 
\hline
\end{tabular}
\caption{List of parameters chosen for the best fit with the real data in few states within India.}
\label{tableS2}
\end{adjustwidth}
\end{table*}

\end{document}